\documentclass[letterpaper]{article} 
\usepackage{aaai25}  
\usepackage{times}  
\usepackage{helvet}  
\usepackage{courier}  
\usepackage[hyphens]{url}  
\usepackage{graphicx} 
\usepackage{natbib}  
\usepackage{caption} 
\usepackage{algorithm}
\usepackage{algorithmic}
\usepackage[table]{xcolor}  
\usepackage[utf8]{inputenc} 
\usepackage[T1]{fontenc}    
\usepackage{url}            
\usepackage{booktabs}       
\usepackage{amsfonts}       
\usepackage{nicefrac}       
\usepackage{microtype}      
\usepackage{xcolor}         
\usepackage{makecell}
\usepackage{adjustbox}
\usepackage{mathtools}
\usepackage{amsmath}
\usepackage{caption}
\usepackage{subcaption}
\usepackage{lipsum} 
\usepackage{dsfont}  
\usepackage{newfloat}
\usepackage{listings}
\DeclareCaptionStyle{ruled}{labelfont=normalfont,labelsep=colon,strut=off} 
\floatstyle{ruled}
\newfloat{listing}{tb}{lst}{}
\floatname{listing}{Listing}
\title{Accelerating High-Fidelity Waveform Generation \\ via Adversarial Flow Matching Optimization}

\author{%
    Sang-Hoon Lee$^{1,2}$,
    Ha-Yeong Choi$^{3}$,
    Seong-Whan Lee$^{4}$
}
 
\affiliations{
    $^{1}$Department of Software and Computer Engineering, Ajou University, Suwon, Korea\\
    $^{2}$Department of Artificial Intelligence, Ajou University, Suwon, Korea\\
    $^{3}$AI Tech Lab, KT Corp., Seoul, Korea\\
    $^{4}$Department of Artificial Intelligence, Korea University, Seoul, Korea
}

\usepackage{bibentry}

\begin{document}

\maketitle

\begin{abstract}
This paper introduces PeriodWave-Turbo, a high-fidelity and high-efficient waveform generation model via adversarial flow matching optimization.
Recently, conditional flow matching (CFM) generative models have been successfully adopted for waveform generation tasks, leveraging a single vector field estimation objective for training. Although these models can generate high-fidelity waveform signals, they require significantly more ODE steps compared to GAN-based models, which only need a single generation step. Additionally, the generated samples often lack high-frequency information due to noisy vector field estimation, which fails to ensure high-frequency reproduction. To address this limitation, we enhance pre-trained CFM-based generative models by incorporating a fixed-step generator modification. We utilized reconstruction losses and adversarial feedback to accelerate high-fidelity waveform generation. Through adversarial flow matching optimization, it only requires 1,000 steps of fine-tuning to achieve state-of-the-art performance across various objective metrics. Moreover, we significantly reduce inference speed from 16 steps to 2 or 4 steps. Additionally, by scaling up the backbone of PeriodWave from 29M to 70M parameters for improved generalization, PeriodWave-Turbo achieves unprecedented performance, with a perceptual evaluation of speech quality (PESQ) score of 4.454 on the LibriTTS dataset. Audio samples, source code and checkpoints will be available at \url{https://github.com/sh-lee-prml/PeriodWave}. 
\end{abstract}

%

\section{Introduction}
This paper explores the methods to accelerate neural ODE waveform generator novel by combining recent advanced generative models, conditional flow matching (CFM) \citep{lipman2022flow,tong2023conditional} with generative adversarial networks (GANs) \citep{goodfellow2014generative,lee2021multi}. Recently, PeriodWave \citep{lee2024periodwave} successfully adopted CFM for neural waveform generation tasks by proposing novel period-aware generator architecture that can reflect the implicit periodic features of waveform signal, and they firstly outperformed the GAN-based neural waveform generation models \citep{kong2020hifi,lee2023bigvgan,shibuya2024bigvsan} which previously dominated the waveform generation tasks. 

However, although PeriodWave could generate high-fidelity waveform signal with iterative refinement, they still require significantly more sampling steps compared to GAN-based models that only need a single generation step. To address this limitation, we propose efficient tuning method, adversarial flow matching optimization which can transform the pre-trained CFM-based generative models into fixed-step generator by leveraging reconstruction loss and adversarial feedback. This simple modification yields state-of-the-art waveform generator by achieving unprecedented performance, with a perceptual evaluation of speech quality (PESQ) \citep{941023} score of 4.454 on LibriTTS benchmark.

In this paper, we propose PeriodWave-Turbo, an efficient but powerful few-steps ODE waveform generator tuned from the pre-trained CFM-based Waveform generator via adversarial flow matching optimization. The main contributions of this work are as follows:

\begin{itemize}
\item We propose PeriodWave-Turbo, a novel ODE-based waveform generator achieving state-of-the-art performance on the objective and subjective evaluation.
\item We successfully accelerate the CFM-based models by adversarial flow matching optimization that utilizes few-step generation and adversarial feedback. 
\item The results shows that our models have much more powerful performance on the two-stage TTS pipelines than other GAN-based models and the pre-trained CFM generator.
\item We demonstrate the effectiveness of the proposed methods by exploring different model sizes. We observed that scaling up the model size simply improved the performance by achieving a PESQ score of 4.454.
\item These improvements only requires few-steps of fine-tuning from the pre-trained CFM teacher models. The entire training time significantly reduced.  
\item  We will release all source code and checkpoints. 
\end{itemize} 
\begin{figure*}[t]
    \centering
    {\includegraphics[width=1\textwidth]{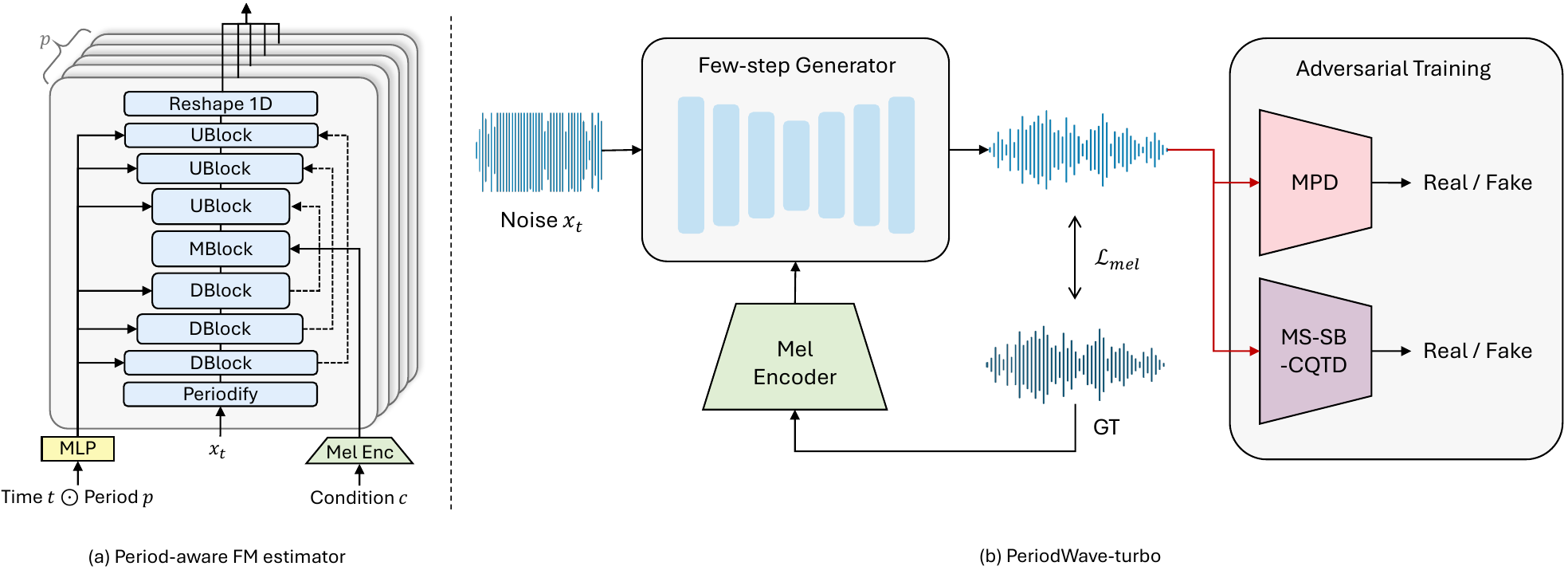}}
    \caption{Overall architiecture of PeriodWave-Turbo. We initialize the parameter of PeriodWave-Turbo by the pre-trained PeriodWave which was trained with flow matching objective. Then, PeriodWave-Turbo is modified by few-step generator with fixed steps. PeriodWave-Turbo is trained with reconstruction Loss and adversarial feedback. Compared to fully GAN training, this could accelerate the training time about $6\times$ faster even with much better performance.}
\end{figure*}

\section{Related Works}\label{section:Preliminary}
\subsection{Accelerating Methods for Few-Step Generator}
Diffusion-based generative models have demonstrated their powerful performance across various domains. However, they require iterative sampling processes, resulting in slow inference speed. To address this limitation, several methods have been proposed to accelerate synthesis speed. 

First, consistency models (CM) \citep{pmlr-v202-song23a,ye2023comospeech} have introduced one-step or few-step generation method by directly mapping noise to data, achieving improved performance through distilling pre-trained models. Consistency trajectory models (CTM) integrated CM and score-based models, and utilized adversarial training to enhance performance. FlashSpeech \citep{ye2024flashspeech} proposed adversarial consistency training by using the SSL-based pre-trained model. 

Denoising diffusion GAN (DDGAN) \citep{xiao2022tackling,oh2024diffprosody} integrated the denoising process with a multimodal conditional GAN to enable faster sampling. UFOGen \citep{xu2024ufogen} enhanced the performance by using an improved reconstruction loss and initializing pre-trained diffusion models for both generator and discriminator. Adversarial diffusion distillation (ADD) \citep{sauer2023adversarial} leveraged a pre-trained teacher model for distillation and utilized adversarial training on the student models for one-step generation. Latent adversarial diffusion distillation (LADD) \citep{sauer2024fast} unified the discriminator and teacher models for more efficient training. Distribution matching distillation (DMD) \citep{yin2024one} demonstrated high-quality one-step generation via distribution matching distillation and reconstruction loss. DMD2 \citep{yin2024improved} significantly improved few-step generation performance by eliminating reconstruction loss and integrating adversarial training with distribution matching distillation. 

In summary, many works demonstrated that a well-designed reconstruction loss and adversarial training are crucial for efficient few-step generator.

\subsection{Adversarial Feedback for Waveform Generation}
So far, GAN-based models have dominated the waveform generation tasks by utilizing various well-designed discriminators to capture the specific characteristics of waveform signals. MelGAN \citep{kumar2019melgan} first proposed a multi-scale discriminator (MSD) that reflects features from different scales of waveform signals using down-sampling. HiFi-GAN \citep{kong2020hifi} introduced the multi-period discriminator (MPD) that can capture the implicit period features by reshaping the waveform signal with specific periodd of prime numbers. UnivNet \citep{jang21_interspeech} presented the multi-resolutional spectrogram discriminator (MRD) to capture different features in spectral domains. Fre-GAN \citep{kim21f_interspeech,lee2022fre} utilized  resolution-wise discriminators. Abocodo \citep{bak2023avocodo} proposed collaborative multi-band discriminator (CoMBD) and sub-band discriminator (SBD). EnCodec \citep{dfossez2023high} modified the MRD by using complex values of spectral features. \citep{gu2024multi} proposed a multi-scale sub-band Constant-Q Transform discriminator (MS-SB-CQTD) that improves the modeling of pitch and harmonic information in waveform signals.

\section{PeriodWave-Turbo}
\subsection{Flow Matching for Waveform Generation}
Recently, the flow matching technique has shown great potential in generating high-quality waveforms by aligning the probability flow between the noise and the target distribution. PeriodWave \citep{lee2024periodwave} employs CFM to create a waveform generator and incorporates a period-aware generator architecture to capture the temporal features of input signals with greater precision. This approach enables the model to learn complex patterns across signals with varying periodicities by extracting and combining features from each period to generate vector fields.  However, despite the effectiveness of flow matching-based models, the iterative processing steps can be slow, posing challenges for real-time applications. Thus, further optimization is necessary to improve generation speed while maintaining output quality.
 
\subsection{Adversarial Flow Matching Optimization}

\paragraph{Few-step Generator Modification}
We accelerate waveform generation by modifying the pre-trained CFM generator into fixed-step generator. We initialize the parameter by the pre-trained PeriodWave, and generate raw waveform signal from the noise $x_0$ with a fixed few-step ODE sampling. We used two or four step generation with Euler method. While this modification restricts the model with fixed step, fine-tuning the model with fixed-step makes it as a specialist for better optimization. Additionally, compared to a single step GAN generator, few-step generator could refine the waveform with iterative samplings \citep{koizumi2023wavefit,jang23b_interspeech,huang2022fastdiff,koizumi22_interspeech,roman2023from} by reducing the train-inference mismatch \citep{tan2024regeneration}. 
However, it is important to fine-tune the model with the properly designed objectives. To do this, we firstly do a deep dive on the acceleration methods for waveform generation by comparing several reconstruction losses, adversarial training, and distillation methods 

\paragraph{Reconstruction Loss} Unlike the pre-training method with flow matching objective, we can utilize reconstruction losses on the raw waveform signals reconstructed by fixed step sampling and ODE solver. According to previous waveform generation models, we adopt the Mel-spectrogram reconstruction loss by transforming the waveform signal by short-time Fourier transformation (STFT) and Mel-filter $\psi$ for focusing the human-perceptual frequency as follows:  
\begin{equation}
\label{Mel_loss}
   L_{mel} = \lVert \psi(x)-\psi(\hat{x})\rVert_1.
\end{equation}
where $\hat{x}$ is sampled with $G(x_t, c, t)$ and ODE solver (Euler method) with fixed steps, c denotes Mel-spectrogram, and [0, 0.25, 0.5, 0.75] for t of four-step generator. 

Furthermore, we adopt multi-scale Mel-spectrogram reconstruction loss by using different parameters of STFT and Mel-filters. Following \citep{kumar2024high}, we also utilizes the hop sizes of [8,16,32,64,128,256,512] and window sizes set to hop\_size$\times$4 to capture the different resolution of frequency information.
For comparison, we conduct the ablation study with multi-scale STFT loss without Mel-filters. But, this decreases the performance by generating high-noisy electric sound on the samples.\footnote{We analyze that phase-level reconstruction loss results in inaccurate phase reproduction of waveform signal. Instead of this, we only adopt multi-scale Mel-spectrogram reconstruction loss for accurate frequency reconstruction.} 
\paragraph{Adversarial Training}
Although multi-scale Mel-spectrogram reconstruction loss could improve the reproduction, stability and convergence speed, it does not guarantee the high-quality waveform generation in subjective perceptual quality by resulting in noisy sound of generated samples. To overcome this issue, we also utilize the adversarial feedback by adopting multi-period discriminator (MPD) and multi-scale sub-band Constant-Q Transform discriminator (MS-SB-CQTD) as follows:
\begin{equation}
  \mathcal{L}_{adv}(D) = \mathbb{E}_{(x)}\Big[(D(x)-1)^2 + D(G(x_t,c,t))^2 \Big],
\end{equation}
\begin{equation}
  \mathcal{L}_{adv}(G) = \mathbb{E}_{x}\Big[(D(G(x_t,c,t))-1)^2 \Big]
\end{equation}
where D consists of MPD and MS-SB-CQTD, and we additionally utilize the feature matching loss $\mathcal{L}_{fm}$ which is the L1 distance between the features of discriminator from the ground-truth $x$ and generated $\hat{x}$. Specifically, MPD can reflect the different features from the different patterns of waveform signal. MS-SB-CQTD can capture the details of frequency information on the multiple scale of CQT spectrogram. Furthermore, additional sub-band processing for different octaves could improve the robustness of pitch modeling. Moreover, BigVGAN-v2 \citep{lee2023bigvgan} demonstrated the effectiveness about the combination of MPD and MS-SB-CQTD. 
\paragraph{Distillation Method} For few-step text-to-image generation, recent works \citep{yin2024one,yin2024improved,sauer2023adversarial,sauer2024fast} introduced a distillation method by using the pre-trained diffusion models. We also compare the distillation method for waveform generation by using the pre-trained FM generator. We utilized the pre-trained PeriodWave as a fake vector field estimator similar to fake score function of \citep{yin2024one}. However, we found that neural vocoder already utilize the time-aligned strong condition Mel-spectrogram so this does not affect the reconstruction performance.\footnote{We see that this distribution matching distillation method could improve the text-conditional audio generation task.}   

\paragraph{Final Loss} The total loss for PeriodWave-Turbo can be expressed as follows:
\begin{equation}
\label{e9}
  \mathcal{L}_{final} = \mathcal{L}_{adv}(G)+\lambda_{fm}\mathcal{L}_{fm}+\lambda_{mel}\mathcal{L}_{mel}
\end{equation}   
where $\lambda_{fm}$ and $\lambda_{mel}$ are set to 2 and 45, respectively, and we utilize four-step generation with Euler method.
\begin{table}[t]
 \caption{Model Details}  \label{params}
  \centering
      \resizebox{1\columnwidth}{!}{
  \begin{tabular}{l|ccc|cc}
    \toprule
    Model & Hidden Dim. & Final Dim. & Params. &Pre-train&Train\\
    \midrule
PeriodWave-S & 256& 16 & 7.57M &2.0d&4.0d\\
PeriodWave-B & 512&32&29.80M&3.0d&4.5d\\
PeriodWave-L & 768& 48&70.24M&5.0d&5.0d\\
    \bottomrule
  \end{tabular}
  } \vspace{-0.4cm}
\end{table}
\subsection{Model Size} We train the PeriodWave and PeriodWave-Turbo with different model sizes. Specifically, we train three models: Small (S, 7.57M), Base (B, 29.80M, original setting), and Large (L, 70.24M) models as indicated in Table \ref{params}.  

\section{Experiment and Result}
\paragraph{Dataset} We train the model with the open-source benchmark dataset, LJSpeech \citep{ljspeech17} and LibriTTS \citep{zen19_interspeech}. Both datasets are widely used for waveform reconstruction tasks. First, LJSpeech is a high-quality single-speaker speech dataset consisting of 13,100 samples with a sampling rate of 22.05 kHz. Following \citep{kong2020hifi}, we utilize the same hyperparameter of STFT and Mel-filters with a hop size of 256, window size of 1024, FFT size of 1024, Mel filter of 80 Bins. Second, LibriTTS dataset is a high-quality multi-speaker speech dataset consisting of 354,780 samples (555 Hours and 2,311 Speakers) with a sampling rate of 24 kHz. Following \citep{lee2023bigvgan}, we utilize the same hyperparameter of STFT and Mel-filters with a hop size of 256, window size of 1024, FFT size of 1024, Mel filter of 100 Bins.

\begin{table*}[t]
 \caption{Objective evaluation results on LJSpeech. We utilized the official checkpoints for all models.}  \label{table1:LJSpeech} 
  \centering
      \resizebox{1\textwidth}{!}{
  \begin{tabular}{l|c|c|ccccc|c}
    \toprule
    Method & Training Steps& Params (M) & M-STFT ($\downarrow$) & PESQ ($\uparrow$) & Period. ($\downarrow$)& V/UV ($\uparrow$) & Pitch ($\downarrow$) & UTMOS ($\uparrow$)   \\
    \midrule
Ground Truth &  - & - & - & - & -& - & -& 4.3804\\
\midrule
HiFi-GAN (V1) & 2.5M& 14.01& 1.0341 & 3.646  & 0.1064 & 0.9584 & 26.839& 4.2691 \\
BigVGAN-base & 5.0M& 14.01& 1.0046 & 3.868  & 0.1054 & 0.9597 & 25.142&4.1986 \\
BigVGAN& 5.0M& 112.4& 0.9369 & 4.210  & 0.0782 & 0.9713 & 19.019& 4.2172 \\
BigVGAN-v2& 3.0M& 112.4& 0.8826 & 4.262  & 0.0663 & 0.9760 & 17.325& 4.3110 \\
\midrule
PriorGrad (50 steps) & 3.0M& 2.61 & 1.2784& 3.918& 0.0879& 0.9661& 17.728& 3.6282\\
\midrule
PeriodWave (16 steps, Midpoint)&1.0M& 29.73& 1.1464 &4.288 & 0.0744 & 0.9704 & 15.042 & 4.3243 \\
PeriodWave+FreeU (16 steps, Midpoint)&1.0M& 29.73& 1.1132 &4.293 & 0.0749 & 0.9701 & 15.753 & 4.3578 \\
\midrule
\rowcolor{gray!15} PeriodWave-Turbo (4 steps, Euler)&1K& 29.73& 0.9253 & 4.349& 0.0640 & 0.9766 & 14.943& 4.3563\\
\rowcolor{gray!15} PeriodWave-Turbo (4 steps, Euler)&5K& 29.73& 0.8969 & 4.344& 0.0588 & 0.9783 & 13.662& 4.3716\\
\rowcolor{gray!15} PeriodWave-Turbo (4 steps, Euler)&10K& 29.73& 0.8880 & 4.387& 0.0624 & 0.9769 & 14.186& 4.3779\\
\rowcolor{gray!15} PeriodWave-Turbo (4 steps, Euler)&0.1M& 29.73& \textbf{0.8352} & \textbf{4.418}& \textbf{0.0572} & \textbf{0.9786} & \textbf{13.473}& \textbf{4.3894}\\
    \bottomrule
  \end{tabular}
  } 
\end{table*}
\begin{table*}[t]
 \caption{Objective evaluation results on the dev subsets of LibriTTS.}  \label{table2:LibriTTS} 
  \centering
      \resizebox{1\textwidth}{!}{
  \begin{tabular}{l|c|c|ccccc|c}
    \toprule
    Method & Training & Params (M) & M-STFT ($\downarrow$) & PESQ ($\uparrow$) & Period. ($\downarrow$)& V/UV ($\uparrow$)& Pitch ($\downarrow$) & UTMOS ($\uparrow$) \\
    \midrule
Ground Truth &  - & - & - & - & - & -& -& 3.8626\\
\midrule
UnivNet-c32 & 1M & 14.87 & 0.8947 & 3.284 & 0.1305&0.9347 & 41.511&3.6085  \\
Vocos & 1M &13.53 & 0.8544 & 3.615&  0.1113& 0.9470 & 33.917& 3.5717 \\
\midrule
BigVSAN & 10M  &112.4  & 0.7210 & 4.316 &  0.0726 & 0.9729 & 21.355 & 3.6921 \\
BigVGAN & 5M  &112.4  & 0.7409 & 4.256 & 0.0809 & 0.9698 & 24.310 & 3.7185\\
BigVGAN-v2 & 3M  &112.4  & \textbf{0.7134} & 4.359 & 0.0621 & \textbf{0.9777} & 21.439 & 3.8362\\
\midrule
 PeriodWave-B (16 steps, Midpoint)& 1M& 29.80 &1.2129& 4.224& 0.0762 & 0.9652 &17.496&3.6495\\
PeriodWave-L (16 steps, Midpoint)& 1M& 70.24& 1.1351& 4.289& 0.0705 & 0.9681 &20.373&3.7139\\
\midrule
 PeriodWave-B + FreeU (16 steps, Midpoint)& 1M& 29.80 &1.0269& 4.248& 0.0765 & 0.9651 &17.398&3.7310\\
PeriodWave-L + FreeU (16 steps, Midpoint)& 1M& 70.24& 0.9788& 4.312& 0.0707 & 0.9682 &19.948&3.7587\\
\midrule
\rowcolor{gray!15} PeriodWave-Turbo-S (4 steps, Euler)& 0.3M& 7.57 & 0.8260& 4.324& 0.0583& 0.9748&\textbf{16.076}&3.8077\\
\rowcolor{gray!15} PeriodWave-Turbo-B (4 steps, Euler)& 0.3M& 29.80 &0.7869& 4.422& 0.0559 & 0.9753 &18.760&3.8450\\
\rowcolor{gray!15} PeriodWave-Turbo-L (4 steps, Euler)&0.3M&70.24&0.7358&\textbf{4.454}&\textbf{0.0528}&0.9758&18.698&\textbf{3.8592}\\
    \bottomrule
  \end{tabular}
  } 
\end{table*}

\paragraph{Pre-training} We pre-train PeriodWave-S/B with flow matching objective for 1M steps using the AdamW optimizer with a learning rate of 2$\times$10$^{-4}$, batch size of 128 on four NVIDIA A100 40GB GPUs. We pre-train PeriodWave-L with the same hyperparameter on four NVIDIA A100 80GB GPUs. During pre-training, we segmented the waveform signal by 32,768 frames for efficient training. For LJSpeech, we only train the Base model of 29.73M. Table \ref{params} indicated the pre-training time of each model for 1M steps. Pre-training the model with FM objective shows tremendous efficiency compared to GAN training. 

\paragraph{Training} To accelerate the CFM-based PeriodWave, we fix the sampling steps with four and use the Euler method for ODE solver. We train the PeriodWave-Turbo from the pre-trained PeriodWave with adversarial flow matching optimization of Equation \ref{e9}. For LJSpeech, we train the PeriodWave-Turbo for only 0.1M steps using the AdamW optimizer with a learning rate of 2$\times$10$^{-5}$, batch size of 32 on four NVIDIA A100 40GB GPUs. For LibriTTS, we train the PeriodWave-Turbo-S/B for 0.3M steps using the AdamW optimizer with a learning rate of 2$\times$10$^{-5}$, batch size of 32 on four NVIDIA A100 40GB GPUs. We train the PeriodWave-Turbo-L with the same hyperparameter on four NVIDIA A100 80GB GPUs. We segmented the waveform signal by 32,768 frames for efficient training. Table \ref{params} indicated the training time of each model for 0.3M steps. Compared to fully GAN training that requires 2.5M \citep{kong2020hifi}, 5M\citep{lee2023bigvgan}, 10M \citep{shibuya2024bigvsan}, our method only requires much smaller training steps (0.3M).

\paragraph{ODE Sampling}
Although the teacher models (PeriodWave) use Midpoint method with sampling steps of 16, PeriodWave-Turbo models utilize Euler method with sampling steps of four. We discussed the inference speed at the Table \ref{table:subject_eval}.   

\subsection{LJSpeech: High-quality Single Speaker Dataset}
We evaluate the LJSpeech benchmark performance by comparing HiFi-GAN, BigVGAN-base, BigVGAN, BigVGAN-v2, PriorGrad, and PeriodWave. Note that BigVGAN-base and BigVGAN are trained with LibriTTS dataset for 5M steps and BigVGAN-v2 is trained with large-scale dataset including multi-lingual speech, environmental sounds, and instruments. Table \ref{table1:LJSpeech} demonstrates that PeriodWave-Turbo achieived state-of-the-art performance in all objective metrics. Furthermore, tuning the model with only 1,000 training steps could achieve better performance than other large-scale trained models without M-STFT. This show the efficiency of our adversarial flow matching optimization. We observed that additional training consistently improve the performance in all metrics. We will release the final checkpoint of PeriodWave-Turbo (LJSpeech) together.
 \begin{table*}[h]
 \caption{Objective evaluation results on the test subsets of LibriTTS.}  \label{table3:LibriTTS}
  \centering
      \resizebox{1\textwidth}{!}{
  \begin{tabular}{l|c|c|ccccc|c}
    \toprule

    Method & Training & Params (M) & M-STFT ($\downarrow$) & PESQ ($\uparrow$) & Period. ($\downarrow$)& V/UV ($\uparrow$)& Pitch ($\downarrow$) & UTMOS ($\uparrow$) \\
    \midrule
Ground Truth &  - & - & - & - & - & -& -& 3.7173\\
\midrule
UnivNet-c32 & 1M & 14.87 & 0.9090 & 3.273 & 0.1293&0.9426 & 53.021&3.5233  \\
Vocos & 1M &13.53 &  0.8703 & 3.647&  0.1008& 0.9584 & 24.032& 3.4916 \\
\midrule
BigVSAN & 10M  &112.4  & \textbf{0.7343} & 4.323 &  0.0650 & 0.9759 & 21.796 & 3.5766 \\
BigVGAN & 5M  &112.4  & 0.7844 & 4.264 & 0.0762 & 0.9692 & 25.651 & 3.5973\\
BigVGAN-v2 & 3M  &112.4  & 0.7400 & 4.266 & 0.0593 & 0.9775 & 19.000 & 3.6948\\
\midrule
 PeriodWave-B (16 steps, Midpoint)& 1M& 29.80 &1.2832& 4.191& 0.0806 & 0.9660 &18.730&3.5325\\
PeriodWave-L (16 steps, Midpoint)& 1M& 70.24& 1.1636& 4.273& 0.0691 & 0.9722 &17.672&3.5862\\
\midrule
 PeriodWave-B + FreeU (16 steps, Midpoint)& 1M& 29.80 &1.0233& 4.220& 0.0742 & 0.9685 &18.495&3.6226\\
PeriodWave-L + FreeU (16 steps, Midpoint)& 1M& 70.24& 0.9929& 4.302& 0.0638 & 0.9739 &17.551&3.6343\\
\midrule
\rowcolor{gray!15} PeriodWave-Turbo-S (4 steps, Euler)& 0.3M& 7.57 &0.8353& 4.324& 0.0566 & 0.9776 &14.020&3.6527\\
\rowcolor{gray!15} PeriodWave-Turbo-B (4 steps, Euler)& 0.3M& 29.80 &0.7963& 4.416& 0.0550 & 0.9784 &16.384&3.7173\\
\rowcolor{gray!15} PeriodWave-Turbo-L (4 steps, Euler)&0.3M&70.24&0.7494&\textbf{4.443}&\textbf{0.0469}&\textbf{0.9816}&\textbf{13.106}&\textbf{3.7229}\\
    \bottomrule
  \end{tabular}
  } 
\end{table*}
\begin{table*}[h]
 \caption{Inference speed and subjective evaluation results on the test subsets of LibriTTS. }  \label{table:subject_eval}
  \centering
      \resizebox{1\textwidth}{!}{
  \begin{tabular}{l|c|ccc|c}
    \toprule 
    Method  & Params (M)&   xRT on CPU ($\uparrow$) &  xRT on GPU ($\uparrow$) & Avg. Memory ($\downarrow$)  & MOS ($\uparrow$)   \\
    \midrule
    GT & - & -&- &- & 4.28$\pm$0.02\\
    \midrule
UnivNet-c32 & 14.87 & $\times$13.53& $\times$267.21 & 1208MB &  4.20$\pm$0.02 \\
Vocos  &13.53 & $\times$85.31 & $\times$436.37 & 133MB &  4.16$\pm$0.02 \\
\midrule
BigVSAN &112.4  & $\times$0.25 & $\times$61.03& 1349MB &  4.21$\pm$0.02 \\
BigVGAN  &112.4  & $\times$0.24& $\times$55.90& 1349MB &  4.18$\pm$0.02 \\
BigVGAN-v2   &112.4  &$\times$0.24 & $\times$112.18 & 1349MB&  4.19$\pm$0.02  \\
\midrule
 PeriodWave-B + FreeU (16 steps, Midpoint)&  29.80 & $\times$0.21 & $\times$4.62 & 897MB &  4.21$\pm$0.02  \\
PeriodWave-L + FreeU (16 steps, Midpoint)& 70.24& $\times$0.09 & $\times$2.77 & 1420MB &  4.20$\pm$0.02 \\
\midrule
\rowcolor{gray!15} PeriodWave-Turbo-S (4 steps, Euler)&  7.57 & $\times$2.58 & $\times$42.26& 413MB &  4.22$\pm$0.02\\
\rowcolor{gray!15} PeriodWave-Turbo-B (4 steps, Euler)&  29.80 & $\times$1.12& $\times$35.01 & 870MB &  4.20$\pm$0.02\\
\rowcolor{gray!15} PeriodWave-Turbo-L (4 steps, Euler)&70.24&  $\times$0.52 & $\times$24.77 & 1393MB&  \textbf{4.23}$\pm$\textbf{0.02} \\
    \bottomrule
  \end{tabular}
  } 
\end{table*} 
\subsection{LibriTTS: Multi-speaker Dataset with 24,000 Hz}
Following BigVGAN, we first evaluate the LibriTTS benchmark performance by comparing UnivNet, Vocos, BigVSAN, BigVGAN, BigVGAN-v2, and PeriodWave. Table \ref{table2:LibriTTS} demonstrated the effectiveness of our model in terms of all metrics. Additionally, PeriodWave-Turbo achieves unprecedented performance, with a perceptual evaluation of speech quality (PESQ) score of 4.454 on this benchmark with only LibriTTS dataset. Furthermore, the results for different model sizes show the robustness of our structure in that PeriodWave-Turbo-S (7.57M) also has better performance than BigVSAN (112.4M) and BigVGAN (112.4M). Furthermore, scaling up the model could improve the performance consistently. Although additional training could decrease the distance of STFT domain, our model already shows better perceptual quality so we stop the training in early steps. To further demonstrate this, we conducted additionally evaluation on subjective evaluation, OOD scenarios, and two-stage TTS.

We also evaluated the objective evaluation for the test subsets of LibriTTS. Table \ref{table3:LibriTTS} shows the consistent results with the dev subsets, and our model also has the best performance on all metrics without M-STFT. Note that the loss curve for M-STFT is not converged but we stop the training for training efficiency.
 
\subsection{Subjective Evaluation}
Additionally, we evaluate the subjective evaluation for the same test subsets of LibriTTS. Table \ref{table:subject_eval} demonstrated that our model has better performance than the previous models including large-scale GAN-based models and teacher models (PeriodWave). The results also demonstrated the effectiveness of our proposed method in that PeriodWave-Turbo-S (7.57M) shows better performance than the large-scale GAN-based models with $\times$14.84 smaller model parameters. PeriodWave-Turbo-L achieved the best performance. 

\subsection{Inference Speed and Memory Usage}
We calculated the xRT on CPU and GPU with AMD EPYC 7313 and NVIDIA RTX A6000, respectively.\footnote{For BigVGAN-v2, we utilized the official CUDA kernel.} xRT denotes the synthesis speed relatative to real-time. We utilize the same samples from the test subsets of LibriTTS. We successfully accelerate the inference speech compared to our teacher models. Specifically, thanks to adversarial flow matching optimization, PeriodWave-Turbo-B could improve the performance with $\times$7.57 faster inference speech compared to PeriodWave-B. Additionally, our models without large model requires lower VRAM usages.      

\begin{table*}[t]
 \caption{Objective evaluation results on out-of-distribution samples from MUSDB18-HQ. We evaluated Periodicity, and V/UV F1 on vocal samples. We highlight the best scores, excluding BigVGAN-v2$^\spadesuit$ as it was not considered for OOD evaluation.}  \label{table6:oodobjective}
  \centering
      \resizebox{0.85\textwidth}{!}{
  \begin{tabular}{l|ccccc}
    \toprule
    Method & M-STFT ($\downarrow$) & PESQ ($\uparrow$) & Periodicity ($\downarrow$)& V/UV F1 ($\uparrow$) & Pitch ($\downarrow$) \\
\midrule
UnivNet-c32 & 1.1377 &1.678  & 0.1588 &0.9186 & 85.317 \\
Vocos  & 1.0203 & 2.173 & 0.1305& 0.9454 & 45.631 \\
\midrule
BigVSAN    & \textbf{0.8766}& 3.011 & 0.0892 & 0.9584  & 37.212 \\
BigVGAN    &  0.9062& 2.862  & 0.0959& 0.9501&30.014 \\
BigVGAN-v2$^\spadesuit$    &  0.8623& 3.470  & 0.0840& 0.9649 &33.229 \\

\midrule
PeriodWave (16 steps, Midpoint)  & 1.2702 & 2.959  & 0.1046 &0.9475&\textbf{27.012} \\
 PeriodWave-L (16 steps, Midpoint)  & 1.1938 & 3.142   & 0.1073  & 0.9382 & 29.346 \\
\midrule
 PeriodWave + FreeU (16 steps, Midpoint)  & 1.1923 & 3.062  & 0.0994 &0.9479 & 36.715 \\
PeriodWave-L + FreeU (16 steps, Midpoint)  & 1.0986 & 3.188& 0.1149 & 0.9461 & 36.350 \\
\midrule
\rowcolor{gray!15} PeriodWave-Turbo-S (4 steps, Euler)& 0.9830&2.980&0.0904&0.9543&27.164\\
\rowcolor{gray!15} PeriodWave-Turbo-B (4 steps, Euler)& 0.9426&3.306&0.0878&0.9519&31.228\\
\rowcolor{gray!15} PeriodWave-Turbo-L (4 steps, Euler) &0.8901&\textbf{3.360}&\textbf{0.0817}&\textbf{0.9597}&28.294\\
    \bottomrule
  \end{tabular}
  } 
\end{table*}
\begin{table*}[h]
 \caption{5-scale SMOS results on out-of-distribution samples from MUSDB18-HQ.}  \label{Table7:OODSMOS} 
  \centering
      \resizebox{0.9\textwidth}{!}{
  \begin{tabular}{l|ccccc|c}
    \toprule
    Method  & Vocal  & Drums & Bass & Others & Mixture & Average\\
    \midrule
Ground Truth &4.48$\pm$0.04&4.45$\pm$0.03&4.46$\pm$0.04 &4.46$\pm$0.03 & 4.46$\pm$0.04 & 4.46$\pm$0.01  \\
\midrule
UnivNet-c32  & 4.10$\pm$0.06&4.04$\pm$0.06&3.73$\pm$0.10 & 3.64$\pm$0.09 &3.74$\pm$0.08 & 3.85$\pm$0.03\\
Vocos  &4.15$\pm$0.06&4.24$\pm$0.05& 3.77$\pm$0.09&3.81$\pm$0.08 &3.95$\pm$0.07& 3.98$\pm$0.03\\
BigVSAN &4.31$\pm$0.05&4.31$\pm$0.05& 4.12$\pm$0.07 &4.21$\pm$0.06 &4.22$\pm$0.06& 4.23$\pm$0.02 \\
BigVGAN  &4.30$\pm$0.06&4.34$\pm$0.05& 4.06$\pm$0.07 &4.21$\pm$0.06 &4.24$\pm$0.05& 4.23$\pm$0.02 \\
BigVGAN-v2$^\spadesuit$ &4.34$\pm$0.05&4.34$\pm$0.05&4.34$\pm$0.06&4.30$\pm$0.05&4.31$\pm$0.05 & 4.33$\pm$0.02\\
\midrule
\rowcolor{gray!15} PeriodWave-B + FreeU (16 steps)&4.30$\pm$0.06&4.28$\pm$0.05& 4.09$\pm$0.07 &4.16$\pm$0.05 &4.21$\pm$0.06 & 4.21$\pm$0.02\\
\rowcolor{gray!15} PeriodWave-L + FreeU (16 steps)& 4.31$\pm$0.06&4.29$\pm$0.05&4.17$\pm$0.07 &4.25$\pm$0.06 &4.20$\pm$0.06& 4.24$\pm$0.02 \\
\midrule
\rowcolor{gray!15} PeriodWave-Turbo-S (4 steps, Euler)& 4.28$\pm$0.06 &4.34$\pm$0.05&4.18$\pm$0.06 &4.23$\pm$0.05&4.20$\pm$0.05& 4.25$\pm$0.02 \\
\rowcolor{gray!15} PeriodWave-Turbo-B (4 steps, Euler)& \textbf{4.34}$\pm$\textbf{0.05}&4.30$\pm$0.05&4.20$\pm$0.06 &4.27$\pm$0.06 & 4.29$\pm$0.05 & 4.28$\pm$0.02\\
\rowcolor{gray!15} PeriodWave-Turbo-L (4 steps, Euler)& 4.30$\pm$0.05&\textbf{4.35}$\pm$\textbf{0.05}&\textbf{4.22}$\pm$\textbf{0.06} &\textbf{4.29}$\pm$\textbf{0.05}&\textbf{4.32}$\pm$\textbf{0.05} & \textbf{4.30}$\pm$\textbf{0.02} \\
    \bottomrule
  \end{tabular}
  }   
\end{table*}

\newpage
\subsection{MUSDB18-HQ: Multi-track Music Audio Dataset for Out-Of-Distribution Robustness}
We conducted the objective evaluation on the MUSDB18-HQ dataset, a multi-track music audio dataset to evaluate the out-of-distribution (OOD) robustness. We compared the performance of our models with the UnivNet, Vocos, BigVSAN, BigVGAN, and BigVGAN-v2.\footnote{BigVGAN-v2$^\spadesuit$ was trained with large-scale dataset including environmental sounds and instruments.} Table \ref{table6:oodobjective} demonstrated the robustness of our model on the OOD samples. we our model performed better at almost objective metrics. 

Additionally, we conducted the subjective evaluation on the MUSDB18-HQ dataset. We compared the similarity MOS for each instruments, and our models has higher SMOS than other models except for BigVGAN-v2 as indicated in Table \ref{Table7:OODSMOS}. We analyzed that using the powerful and novel period-aware vector field estimator for our generator could refine the waveform signal with iterative samplings, and the Table \ref{table6:oodobjective} also demonstrated that the effectiveness of pitch and periodicity modeling resulting in better audio quality and similarity. For model scalability, scaling up the model could improve the robustness of out-of-distribution scenarios. 

\begin{table*}
\caption{Zero-shot TTS Results. We utilized ARDiT-TTS trained with LibriTTS as TTS model.}
\label{TTSMOS}
\centering
\resizebox{1\textwidth}{!}{
\begin{tabular}{l|ccccccc}
\toprule
Methods & CER ($\downarrow$) & WER ($\downarrow$) & WavLM ($\uparrow$) & Resemblyzer ($\uparrow$) &  UTMOS ($\uparrow$)   & MOS ($\uparrow$) \\
\midrule
BigVSAN         & 0.58 & 1.63  & 0.573 & 0.835 &  3.9732 &  4.26$\pm$0.02\\
BigVGAN         & 0.59 & 1.60  & 0.576 & 0.836 &  4.0424  & 4.28$\pm$0.02\\
BigVGAN-v2      & 0.54 & 1.53  & \textbf{0.579} &0.835 & 3.9894 & 4.24$\pm$0.02 \\
\midrule
 PeriodWave-B (16 steps, Midpoint)  & 0.59 & 1.65 &0.553& \textbf{0.837}  & 4.2209 &  4.30$\pm$0.02 \\
PeriodWave-L (16 steps, Midpoint) & 0.59 & 1.65 & 0.557&0.821  & 4.2464 &  4.27$\pm$0.02 \\
\midrule
 PeriodWave-B + FreeU (16 steps, Midpoint) & 0.59 & 1.65 & 0.556 &0.835  & 4.2621 &  4.31$\pm$0.02  \\
PeriodWave-L + FreeU (16 steps, Midpoint)& 0.57 & 1.60 &0.560 & 0.833  &   4.2643 &  4.29$\pm$0.02  \\
\midrule
\rowcolor{gray!15} PeriodWave-Turbo-S (4 steps, Euler)    & 0.64 & 1.64 & 0.575& 0.835& 4.2716 &  \textbf{4.34}$\pm$\textbf{0.02}   \\
\rowcolor{gray!15} PeriodWave-Turbo-B (4 steps, Euler)    & 0.52 & 1.52 & 0.576&0.835  & 4.2932 &  4.33$\pm$0.02  \\
\rowcolor{gray!15} PeriodWave-Turbo-L (4 steps, Euler)  & \textbf{0.50} & \textbf{1.49} & \textbf{0.579}& 0.835  & \textbf{4.3026} &  4.32$\pm$0.02   \\
\bottomrule
\end{tabular}
}
\end{table*}
\begin{table*}[t]
 \caption{Ablation Study results on the dev subsets of LibriTTS.}  \label{table9:ablation}
  \centering
      \resizebox{1\textwidth}{!}{
  \begin{tabular}{l|c|ccccc|c}
    \toprule
    Method & Training Steps & M-STFT ($\downarrow$) & PESQ ($\uparrow$) & Periodicity ($\downarrow$)& V/UV F1 ($\uparrow$)& Pitch ($\downarrow$) & UTMOS ($\uparrow$) \\
    \midrule
    PeriodWave-Turbo-B (2 steps, Euler)& 0.3M &0.8353& 4.345& 0.0614 & 0.9702 &16.460&3.8089\\
        \midrule
\rowcolor{gray!15} PeriodWave-Turbo-B (4 steps, Euler) & 0.3M&0.7869& 4.422& 0.0559 & 0.9753 &18.760&3.8450\\
- GAN & 10K &0.8267&4.356&0.0627&0.9709&19.361&3.8084\\
- GAN - M-Mel Loss + Single Mel Loss & 10K&0.8198&4.306&0.0709&0.9661&19.939&2.7801 \\
- GAN - M-Mel Loss + M-STFT Loss & 10K & 0.8286& 4.297&0.0686&0.9691&16.688&2.7695\\

\midrule
- GAN + DMD & 10K&0.8407&4.362&0.0615&0.9721&15.888&3.7990 \\
\midrule
from the scratch (- pre-training with FM) & 0.3M & 0.9005&3.825&0.1109&0.9459&36.209&3.7464\\
from the scratch (- pre-training with FM) & 0.5M & 0.8811&3.935&0.1014&0.9501&29.353&3.7898\\
    \bottomrule
  \end{tabular}
  } 
\end{table*}

\subsection{Zero-shot TTS Results on LibriTTS dataset}
Previously, large-scale GAN-based models show low performance on two-stage TTS which consists of acoustic model and neural vocoder due to training-inference mismatch issue. Although they reconstruct the waveform signal from the ground-truth Mel-spectrogram, they might lose their robustness on the synthesized Mel-spectrogram by acoustic models due to the imperfect conditioning \citep{roman2023from}. 

To further demonstrate the effectiveness of our models, we conduct the evaluation on two-stage TTS pipelines with ARDiT-TTS \citep{liu2024autoregressive}. We calculate the character error rate (CER) and word error rate (WER) by using Whishper-large-v2, and speaker encoder cosine similarity (SECS) by using WavLM and Resemblyzer, UTMOS for audio quality, and subjective evaluation MOS.

Table \ref{TTSMOS} shows that PeriodWave-Turbo-B/L have better speech accuracy in terms of CER and WER than other previous models. All models have almost same SECS results, and this means that SECS is more related to acoustic model. The UTMOS results show that our models have better audio quality. The subjective evaluation results also demonstrate our model have better naturalness on two-stage TTS scenarios.

\subsection{Ablation Study}
\paragraph{Reconstruction Loss} We compared three reconstruction losses including multi-scale Mel-spectrogram (M-Mel) loss, single-scale Mel-spectrogram (Single Mel) loss, and multi-scale STFT (M-STFT) loss at Table \ref{table9:ablation}. We found that most models without GAN collapsed after 10k steps and only M-Mel loss could make the training stable without GAN loss. Single Mel loss has lower performance than M-Mel loss, and the model with M-STFT generates a high electric noise. 

\paragraph{Adversarial Feedback} Although M-Mel loss could improve the robustness in terms of objective metrics, the generated samples contain a lot of artifact resulting in low perceptual quality. To reduce this problem, it is crucial to utilize the adversarial feedback with well designed discriminator. With MPD and MS-SB-CQTD, our model can be trained stably, and the performance consistently improved during training.

\paragraph{Distribution Matching Distillation} We found that DMD with the pre-trained FM generator as a teacher model could not improve the performance. We can discuss it because we already use the strong time-aligned condition Mel-spectrogram for neural vocoder so DMD is not necessary. However, we see that adopting this loss could improve the performance of text-conditional generation tasks such as text-to-audio model.

\paragraph{Pre-training} It is important to utilize the pre-trained CFM generator for adversarial flow matching optimization. First, this significantly reduce the entire training time. comparing the models trained with same time, PeriodWave-Turbo-B performed much better than the model from the scratch (0.5M). 

\paragraph{Few-step Generation} We trained the model by modifying it as two-step generator. This model also achieved higher performance than other previous models, and has better efficiency. However, increasing the steps results in better performance so we utilize four-step generation as default settings.

\section{Conclusion}
In this work, we present PeriodWave-Turbo, a novel ODE-based few-step waveform generator. With adversarial flow matching optimization, we successfully accelerate the CFM-based waveform generator. Our results demonstrate both the effectiveness and efficiency of our approach, achieving state-of-the-art performance across nearly all objective and subjective metrics. Notably, our model attained an unprecedented PESQ score of 4.454 on the LibriTTS benchmark. Furthermore, our model exhibits superior robustness in OOD and two-stage TTS scenarios. 

While our models demonstrate powerful generative performance and training efficiency, we recognize the opportunity to further optimize inference speed. In future work, we intend to enhance inference speed by adopting multiple STFT-based down-sampling method, replacing the current downsampling block in UNet. 
Additionally, we believe that our model has the potential to be adapted for end-to-end text-to-speech and text-to-audio generation tasks. Therefore, we plan to extend our work to encompass versatile conditional generative models. We hope that our approach will contribute to the study of waveform generation, and as such, we will release all source code, checkpoints, and TTS/VC application \citep{choi2023diff,choi2024dddm} together.

\newpage
 \section*{Acknowledgement} \label{section:ack}
We sincerely thank the authors of HiFi-GAN and BigVGAN for their dedication in making their groundbreaking work open-source, greatly benefiting the speech research community. We sincerely thank Zhijun Liu, the author of ARDiT-TTS for providing the Mel-spectrograms of ARDiT-TTS, which enabled us to perform the second stage of TTS synthesis. 

This work was supported by Institute of Information \& communications Technology Planning \& Evaluation (IITP) grant funded by the Korea government (MSIT) (IITP-2024-RS-2023-00255968, the Artificial Intelligence Convergence Innovation Human Resources Development and No. 2021-0-02068, Artificial Intelligence Innovation Hub) and Artificial intelligence industrial convergence cluster development project funded by the Ministry of Science and ICT(MSIT, Korea)\&Gwangju Metropolitan City.

\bibliography{aaai25}

\begin{thebibliography}{38}
\providecommand{\natexlab}[1]{#1}

\bibitem[{Bak et~al.(2023)Bak, Lee, Bae, Yang, Bae, and Joo}]{bak2023avocodo}
Bak, T.; Lee, J.; Bae, H.; Yang, J.; Bae, J.-S.; and Joo, Y.-S. 2023.
\newblock Avocodo: Generative adversarial network for artifact-free vocoder.
\newblock In \emph{Proceedings of the AAAI Conference on Artificial Intelligence}, volume~37, 12562--12570.

\bibitem[{Choi, Lee, and Lee(2023)}]{choi2023diff}
Choi, H.-Y.; Lee, S.-H.; and Lee, S.-W. 2023.
\newblock Diff-HierVC: Diffusion-based hierarchical voice conversion with robust pitch generation and masked prior for zero-shot speaker adaptation.
\newblock \emph{International Speech Communication Association}, 2283--2287.

\bibitem[{Choi, Lee, and Lee(2024)}]{choi2024dddm}
Choi, H.-Y.; Lee, S.-H.; and Lee, S.-W. 2024.
\newblock Dddm-vc: Decoupled denoising diffusion models with disentangled representation and prior mixup for verified robust voice conversion.
\newblock In \emph{Proceedings of the AAAI Conference on Artificial Intelligence}, volume~38, 17862--17870.

\bibitem[{D{\'e}fossez et~al.(2023)D{\'e}fossez, Copet, Synnaeve, and Adi}]{dfossez2023high}
D{\'e}fossez, A.; Copet, J.; Synnaeve, G.; and Adi, Y. 2023.
\newblock High Fidelity Neural Audio Compression.
\newblock \emph{Transactions on Machine Learning Research}.
\newblock Featured Certification, Reproducibility Certification.

\bibitem[{Goodfellow et~al.(2014)Goodfellow, Pouget-Abadie, Mirza, Xu, Warde-Farley, Ozair, Courville, and Bengio}]{goodfellow2014generative}
Goodfellow, I.; Pouget-Abadie, J.; Mirza, M.; Xu, B.; Warde-Farley, D.; Ozair, S.; Courville, A.; and Bengio, Y. 2014.
\newblock Generative adversarial nets.
\newblock \emph{Advances in neural information processing systems}, 27.

\bibitem[{Gu et~al.(2024)Gu, Zhang, Xue, and Wu}]{gu2024multi}
Gu, Y.; Zhang, X.; Xue, L.; and Wu, Z. 2024.
\newblock Multi-scale sub-band constant-q transform discriminator for high-fidelity vocoder.
\newblock In \emph{ICASSP 2024-2024 IEEE International Conference on Acoustics, Speech and Signal Processing (ICASSP)}, 10616--10620. IEEE.

\bibitem[{Huang et~al.(2022)Huang, Lam, Wang, Su, Yu, Ren, and Zhao}]{huang2022fastdiff}
Huang, R.; Lam, M.~W.; Wang, J.; Su, D.; Yu, D.; Ren, Y.; and Zhao, Z. 2022.
\newblock Fastdiff: A fast conditional diffusion model for high-quality speech synthesis.
\newblock \emph{arXiv preprint arXiv:2204.09934}.

\bibitem[{Ito and Johnson(2017)}]{ljspeech17}
Ito, K.; and Johnson, L. 2017.
\newblock The LJ Speech Dataset.
\newblock \url{https://keithito.com/LJ-Speech-Dataset/}.

\bibitem[{Jang, Lim, and Park(2023)}]{jang23b_interspeech}
Jang, W.; Lim, D.; and Park, H. 2023.
\newblock {FastFit: Towards Real-Time Iterative Neural Vocoder by Replacing U-Net Encoder With Multiple STFTs}.
\newblock In \emph{Proc. INTERSPEECH 2023}, 4364--4368.

\bibitem[{Jang et~al.(2021)Jang, Lim, Yoon, Kim, and Kim}]{jang21_interspeech}
Jang, W.; Lim, D.; Yoon, J.; Kim, B.; and Kim, J. 2021.
\newblock {UnivNet: A Neural Vocoder with Multi-Resolution Spectrogram Discriminators for High-Fidelity Waveform Generation}.
\newblock In \emph{Proc. Interspeech 2021}, 2207--2211.

\bibitem[{Kim et~al.(2021)Kim, Lee, Lee, and Lee}]{kim21f_interspeech}
Kim, J.-H.; Lee, S.-H.; Lee, J.-H.; and Lee, S.-W. 2021.
\newblock {Fre-GAN: Adversarial Frequency-Consistent Audio Synthesis}.
\newblock In \emph{Proc. Interspeech 2021}, 2197--2201.

\bibitem[{Koizumi et~al.(2023)Koizumi, Yatabe, Zen, and Bacchiani}]{koizumi2023wavefit}
Koizumi, Y.; Yatabe, K.; Zen, H.; and Bacchiani, M. 2023.
\newblock WaveFit: An iterative and non-autoregressive neural vocoder based on fixed-point iteration.
\newblock In \emph{2022 IEEE Spoken Language Technology Workshop (SLT)}, 884--891. IEEE.

\bibitem[{Koizumi et~al.(2022)Koizumi, Zen, Yatabe, Chen, and Bacchiani}]{koizumi22_interspeech}
Koizumi, Y.; Zen, H.; Yatabe, K.; Chen, N.; and Bacchiani, M. 2022.
\newblock {SpecGrad: Diffusion Probabilistic Model based Neural Vocoder with Adaptive Noise Spectral Shaping}.
\newblock In \emph{Proc. Interspeech 2022}, 803--807.

\bibitem[{Kong, Kim, and Bae(2020)}]{kong2020hifi}
Kong, J.; Kim, J.; and Bae, J. 2020.
\newblock Hifi-gan: Generative adversarial networks for efficient and high fidelity speech synthesis.
\newblock \emph{Advances in neural information processing systems}, 33: 17022--17033.

\bibitem[{Kumar et~al.(2019)Kumar, Kumar, De~Boissiere, Gestin, Teoh, Sotelo, De~Brebisson, Bengio, and Courville}]{kumar2019melgan}
Kumar, K.; Kumar, R.; De~Boissiere, T.; Gestin, L.; Teoh, W.~Z.; Sotelo, J.; De~Brebisson, A.; Bengio, Y.; and Courville, A.~C. 2019.
\newblock Melgan: Generative adversarial networks for conditional waveform synthesis.
\newblock \emph{Advances in neural information processing systems}, 32.

\bibitem[{Kumar et~al.(2024)Kumar, Seetharaman, Luebs, Kumar, and Kumar}]{kumar2024high}
Kumar, R.; Seetharaman, P.; Luebs, A.; Kumar, I.; and Kumar, K. 2024.
\newblock High-fidelity audio compression with improved rvqgan.
\newblock \emph{Advances in Neural Information Processing Systems}, 36.

\bibitem[{Lee et~al.(2023)Lee, Ping, Ginsburg, Catanzaro, and Yoon}]{lee2023bigvgan}
Lee, S.-g.; Ping, W.; Ginsburg, B.; Catanzaro, B.; and Yoon, S. 2023.
\newblock Big{VGAN}: A Universal Neural Vocoder with Large-Scale Training.
\newblock In \emph{The Eleventh International Conference on Learning Representations}.

\bibitem[{Lee, Choi, and Lee(2024)}]{lee2024periodwave}
Lee, S.-H.; Choi, H.-Y.; and Lee, S.-W. 2024.
\newblock PeriodWave: Multi-Period Flow Matching for High-Fidelity Waveform Generation.
\newblock \emph{arXiv preprint arXiv:2408.07547}.

\bibitem[{Lee et~al.(2022)Lee, Kim, Lee, and Lee}]{lee2022fre}
Lee, S.-H.; Kim, J.-H.; Lee, K.-E.; and Lee, S.-W. 2022.
\newblock Fre-gan 2: Fast and efficient frequency-consistent audio synthesis.
\newblock In \emph{ICASSP 2022-2022 IEEE International Conference on Acoustics, Speech and Signal Processing (ICASSP)}, 6192--6196. IEEE.

\bibitem[{Lee et~al.(2021)Lee, Yoon, Noh, Kim, and Lee}]{lee2021multi}
Lee, S.-H.; Yoon, H.-W.; Noh, H.-R.; Kim, J.-H.; and Lee, S.-W. 2021.
\newblock Multi-spectrogan: High-diversity and high-fidelity spectrogram generation with adversarial style combination for speech synthesis.
\newblock In \emph{Proceedings of the AAAI Conference on Artificial Intelligence}, volume~35, 13198--13206.

\bibitem[{Lipman et~al.(2022)Lipman, Chen, Ben-Hamu, Nickel, and Le}]{lipman2022flow}
Lipman, Y.; Chen, R.~T.; Ben-Hamu, H.; Nickel, M.; and Le, M. 2022.
\newblock Flow matching for generative modeling.
\newblock \emph{arXiv preprint arXiv:2210.02747}.

\bibitem[{Liu et~al.(2024)Liu, Wang, Inoue, Bai, and Li}]{liu2024autoregressive}
Liu, Z.; Wang, S.; Inoue, S.; Bai, Q.; and Li, H. 2024.
\newblock Autoregressive Diffusion Transformer for Text-to-Speech Synthesis.
\newblock \emph{arXiv preprint arXiv:2406.05551}.

\bibitem[{Oh, Lee, and Lee(2024)}]{oh2024diffprosody}
Oh, H.-S.; Lee, S.-H.; and Lee, S.-W. 2024.
\newblock Diffprosody: Diffusion-based latent prosody generation for expressive speech synthesis with prosody conditional adversarial training.
\newblock \emph{IEEE/ACM Transactions on Audio, Speech, and Language Processing}.

\bibitem[{Rix et~al.(2001)Rix, Beerends, Hollier, and Hekstra}]{941023}
Rix, A.; Beerends, J.; Hollier, M.; and Hekstra, A. 2001.
\newblock Perceptual evaluation of speech quality (PESQ)-a new method for speech quality assessment of telephone networks and codecs.
\newblock In \emph{2001 IEEE International Conference on Acoustics, Speech, and Signal Processing. Proceedings (Cat. No.01CH37221)}, volume~2, 749--752 vol.2.

\bibitem[{Roman et~al.(2023)Roman, Adi, Deleforge, Serizel, Synnaeve, and D{\'e}fossez}]{roman2023from}
Roman, R.~S.; Adi, Y.; Deleforge, A.; Serizel, R.; Synnaeve, G.; and D{\'e}fossez, A. 2023.
\newblock From Discrete Tokens to High-Fidelity Audio Using Multi-Band Diffusion.
\newblock In \emph{Thirty-seventh Conference on Neural Information Processing Systems}.

\bibitem[{Sauer et~al.(2024)Sauer, Boesel, Dockhorn, Blattmann, Esser, and Rombach}]{sauer2024fast}
Sauer, A.; Boesel, F.; Dockhorn, T.; Blattmann, A.; Esser, P.; and Rombach, R. 2024.
\newblock Fast high-resolution image synthesis with latent adversarial diffusion distillation.
\newblock \emph{arXiv preprint arXiv:2403.12015}.

\bibitem[{Sauer et~al.(2023)Sauer, Lorenz, Blattmann, and Rombach}]{sauer2023adversarial}
Sauer, A.; Lorenz, D.; Blattmann, A.; and Rombach, R. 2023.
\newblock Adversarial diffusion distillation.
\newblock \emph{arXiv preprint arXiv:2311.17042}.

\bibitem[{Shibuya, Takida, and Mitsufuji(2024)}]{shibuya2024bigvsan}
Shibuya, T.; Takida, Y.; and Mitsufuji, Y. 2024.
\newblock Bigvsan: Enhancing gan-based neural vocoders with slicing adversarial network.
\newblock In \emph{ICASSP 2024-2024 IEEE International Conference on Acoustics, Speech and Signal Processing (ICASSP)}, 10121--10125. IEEE.

\bibitem[{Song et~al.(2023)Song, Dhariwal, Chen, and Sutskever}]{pmlr-v202-song23a}
Song, Y.; Dhariwal, P.; Chen, M.; and Sutskever, I. 2023.
\newblock Consistency Models.
\newblock In \emph{Proceedings of the 40th International Conference on Machine Learning}, volume 202 of \emph{Proceedings of Machine Learning Research}, 32211--32252. PMLR.

\bibitem[{Tan et~al.(2024)Tan, Qin, Bian, Liu, and Bengio}]{tan2024regeneration}
Tan, X.; Qin, T.; Bian, J.; Liu, T.-Y.; and Bengio, Y. 2024.
\newblock Regeneration learning: A learning paradigm for data generation.
\newblock In \emph{Proceedings of the AAAI Conference on Artificial Intelligence}, volume~38, 22614--22622.

\bibitem[{Tong et~al.(2023)Tong, Malkin, Huguet, Zhang, Rector-Brooks, Fatras, Wolf, and Bengio}]{tong2023conditional}
Tong, A.; Malkin, N.; Huguet, G.; Zhang, Y.; Rector-Brooks, J.; Fatras, K.; Wolf, G.; and Bengio, Y. 2023.
\newblock Conditional flow matching: Simulation-free dynamic optimal transport.
\newblock \emph{arXiv preprint arXiv:2302.00482}, 2(3).

\bibitem[{Xiao, Kreis, and Vahdat(2022)}]{xiao2022tackling}
Xiao, Z.; Kreis, K.; and Vahdat, A. 2022.
\newblock Tackling the Generative Learning Trilemma with Denoising Diffusion {GAN}s.
\newblock In \emph{International Conference on Learning Representations}.

\bibitem[{Xu et~al.(2024)Xu, Zhao, Xiao, and Hou}]{xu2024ufogen}
Xu, Y.; Zhao, Y.; Xiao, Z.; and Hou, T. 2024.
\newblock Ufogen: You forward once large scale text-to-image generation via diffusion gans.
\newblock In \emph{Proceedings of the IEEE/CVF Conference on Computer Vision and Pattern Recognition}, 8196--8206.

\bibitem[{Ye et~al.(2024)Ye, Ju, Liu, Tan, Chen, Lu, Sun, Pan, Bianweizhen, He, Xue, Liu, and Guo}]{ye2024flashspeech}
Ye, Z.; Ju, Z.; Liu, H.; Tan, X.; Chen, J.; Lu, Y.; Sun, P.; Pan, J.; Bianweizhen; He, S.; Xue, W.; Liu, Q.; and Guo, Y. 2024.
\newblock FlashSpeech: Efficient Zero-Shot Speech Synthesis.
\newblock In \emph{ACM Multimedia 2024}.

\bibitem[{Ye et~al.(2023)Ye, Xue, Tan, Chen, Liu, and Guo}]{ye2023comospeech}
Ye, Z.; Xue, W.; Tan, X.; Chen, J.; Liu, Q.; and Guo, Y. 2023.
\newblock Comospeech: One-step speech and singing voice synthesis via consistency model.
\newblock In \emph{Proceedings of the 31st ACM International Conference on Multimedia}, 1831--1839.

\bibitem[{Yin et~al.(2024{\natexlab{a}})Yin, Gharbi, Park, Zhang, Shechtman, Durand, and Freeman}]{yin2024improved}
Yin, T.; Gharbi, M.; Park, T.; Zhang, R.; Shechtman, E.; Durand, F.; and Freeman, W.~T. 2024{\natexlab{a}}.
\newblock Improved Distribution Matching Distillation for Fast Image Synthesis.
\newblock \emph{arXiv preprint arXiv:2405.14867}.

\bibitem[{Yin et~al.(2024{\natexlab{b}})Yin, Gharbi, Zhang, Shechtman, Durand, Freeman, and Park}]{yin2024one}
Yin, T.; Gharbi, M.; Zhang, R.; Shechtman, E.; Durand, F.; Freeman, W.~T.; and Park, T. 2024{\natexlab{b}}.
\newblock One-step diffusion with distribution matching distillation.
\newblock In \emph{Proceedings of the IEEE/CVF Conference on Computer Vision and Pattern Recognition}, 6613--6623.

\bibitem[{Zen et~al.(2019)Zen, Dang, Clark, Zhang, Weiss, Jia, Chen, and Wu}]{zen19_interspeech}
Zen, H.; Dang, V.; Clark, R.; Zhang, Y.; Weiss, R.~J.; Jia, Y.; Chen, Z.; and Wu, Y. 2019.
\newblock {LibriTTS: A Corpus Derived from LibriSpeech for Text-to-Speech}.
\newblock In \emph{Proc. Interspeech 2019}, 1526--1530.

\end{thebibliography}

\end{document}